\documentclass[intlimits,twoside,a4paper]{article}
\usepackage{amsmath,amssymb}
\usepackage{graphicx}

\usepackage[T2A]{fontenc}
\usepackage[cp1251]{inputenc}

\usepackage[eqsecnum]{cmpj2}

\issue{2017}{20}{1}{13801}
\doinumber{10.5488/CMP.20.13801}

\title[Negative response to an excessive bias]%
{Negative response to an excessive bias by a mixed population of voters%
\thanks{This paper is dedicated to Yu. Holovatch on the occasion of his 60th anniversary.}}
\author[V.S. Dotsenko, C. Mej{\'i}a-Monasterio, G. Oshanin]{V.S. Dotsenko\refaddr{label1},
        C. Mej{\'i}a-Monasterio\refaddr{label2}, G. Oshanin\refaddr{label1}}
\addresses{
\addr{label1} Laboratoire de Physique Th\'{e}orique de la Mati\`{e}re Condens\'{e}e, UPMC,
CNRS UMR 7600, Sorbonne Universit\'{e}s,\\ 4 Place Jussieu, 75252 Paris Cedex 05, France
\addr{label2} Laboratory of Physical Properties, Technical University of Madrid, Av. Complutense s/n 28040 Madrid, Spain
}

\authorcopyright{V.S. Dotsenko, C. Mej{\'i}a-Monasterio, G. Oshanin, 2017}
\date{Received January 27, 2017, in final form February 28, 2017}

\begin{document}

\maketitle

\begin{abstract}
We study an outcome of a vote in a population of voters exposed to an  externally applied bias in favour of one of two potential candidates. The population consists of ordinary individuals, that are in majority and tend to align their opinion with the external bias, and some number of contrarians --- individuals
who are  always hostile to the bias but are not in a conflict with ordinary voters. The voters interact among themselves, all with all, trying to find an opinion reached by the community  as a whole. We demonstrate that for a sufficiently weak external bias, the opinion of ordinary individuals is always decisive and the outcome of the vote
is in favour of the preferential candidate.
On the contrary, for an excessively strong bias,
the contrarians dominate
in the population's opinion, producing overall a negative response
to the imposed bias.
We also show that for sufficiently strong interactions within the community, either of two subgroups can abruptly change an opinion of the other group.
\keywords non-linear and negative response, external bias, population of voters
\pacs  89.65.Ef, 89.75.-k
\end{abstract}

\section{Introduction}

One often encounters situations in which trying harder,  pushing stronger or, in general, making any excessive  effort appears to be counterproductive and leads to a smaller (or even a negative) effect, as compared to the outcome achieved with a more modest investment. This is so, in particular, for long-lasting human relationships and, on a larger scale, for a cohabitation of different countries. In condensed matter physics, such an effect takes place in many instances. To name but  a few, we mention
electron transfer in semiconductors at low temperatures~\cite{conwell,NCCGO76,SBW86,LHC91},
hopping processes in disordered media \cite{bryksin,barma}, transport of electrons in mixtures of atomic gases \cite{vrh}, some models of Brownian motors~\cite{SGN97,KMHLT06,gleb1,gleb2},  transport in
kinetically constrained models of glass formers~\cite{JKGC08} and in lattice gases \cite{MB0,LF13,BBMS13,MB1,al1,MB2,al2_1,al2_2,al3}, which also exhibits other spectacular anomalies beyond the linear-response regime \cite{1,11,2,3}. In all these physical examples, exerting more force on an object dragged through a system yields a disproportionally strong response of the medium, so that the resulting velocity of the object appears to be less for strong forces than for the moderate ones. Similar phenomena also occur, due to other physical reasons, in far-from-equilibrium quantum spin chains \cite{cas}, systems which exhibit a ``freezing-by-heating'' behavior (see, e.g., \cite{malte})  and
different types of non-equilibrium systems \cite{royce}.

In the present paper we discuss a toy model
of opinion formation in a mixed society
in which applying
an excessive external bias appears to be counterproductive.
We consider here a big society
which
consists of many small communities, each comprising
$N$ individuals. These individuals have to vote for and eventually  to
elect
either of two candidates --- candidate $1$ and candidate $2$.
The individuals within each community
interact between
themselves, all with all,
trying to achieve a
consensus ---
an opinion reached by the group as a whole.
The society lives at some effective temperature $T$, which permits for
fluctuations in opinions
within a given community.

The voters are exposed to an external bias, e.g., due to mass media,
prompting them to choose a preferential candidate.
With respect to the external bias,
each community is divided into two subgroups, of sizes $N_1 = (1- \rho) N$ and $N_2 = N - N_1 = \rho N$ with $0 \leqslant \rho \leqslant 1$, respectively.
The individuals that belong to the first subgroup --- we call them ``ordinary voters'' --- tend to adjust their opinion to the external bias.
The voters that form
the second subgroup --- a sort of  ``contrarians'' in a model of Galam \cite{galam,sen}
--- are opposed to the external bias and tend to have an opinion different from the \textit{externally} imposed one.\footnote{We emphasize that in \cite{galam,sen} the contrarians are at odds with ordinary voters, which represent the majority. In our settings, they rather contradict the external bias but are not in a conflict with ordinary voters.}. The opinions of the ordinary voters and of the contrarians may have a different \textit{strength}.
Indeed, as it usually happens, the contrarians
may be more persuasive, may mobilize themselves more readily and also may
have more influence on the community than the ordinary voters. We are interested to know the eventual
outcome of the elections in the whole society and, more specifically, to get a qualitative, conceptual
understanding of how an intensity
of the external bias affects this outcome.
We proceed to show that upon
exposing the system to a bias and gradually increasing its magnitude,
a series of interesting phenomena may take place, including, e.g., an abrupt change
of the opinion of each of the subgroups when the ``ties'' within a given community are sufficiently strong.
We will also demonstrate that
one first
achieves an opinion aligned with the bias. Increasing the bias further
 up to some critical value, gives a maximal effect with a saturation of the bias-induced opinion, and then, upon
 exceeding this critical value, one observes a decrease of the latter and eventually, a formation of an opinion
 which is antagonistic to the bias.

\section{The model}

To put our toy model into
formal terms, we assign to each of the individuals a ``spin'' variable
characterising the strength of his or her opinion in favour of one of the candidates.
For ordinary individuals, this ``spin'' variable is defined as $s_i$ (with index $i$ labelling these individuals, $i=1, \ldots, N_1$)
and
is rigid
assuming only two values: $s_i = +1$ corresponding to the opinion in favor of
the candidate $1$, and $s_i = - 1$ corresponding to the preference for the candidate $2$.
To describe next
the strength of the
 opinion of each of the contrarians and to take into account their persuasive ability, we introduce a soft ``spin''
variable $\phi_j$ ($j=1, \ldots, N_2$), and suppose that it
can, in principle, attain any real positive or negative
value, but its distribution $P(\phi_j)$ is sharply peaked at $\phi_j=-1$ and $\phi_j =1$.
As an example,
we will consider Gaussian distribution of the form
\begin{align}
\label{P}
P(\phi) = \frac{1}{\sqrt{8 \pi \sigma_{\phi}}} \left\{\exp\left[- \frac{(\phi-1)^2}{2 \sigma_{\phi}}\right]+ \exp\left[- \frac{(\phi+1)^2}{2 \sigma_{\phi}}\right]\right\} \,,
\end{align}
 where $\sigma_{\phi}$ characterises the effective stiffness of the distribution.
 When $\sigma_{\phi} \to 0$,  the distribution of $\phi_j$ in~\eqref{P}
becomes the sum of two delta-functions, just like the one for $s_i$.

Further on, we suppose that our system comprising rigid and soft spins experiences an action of an external bias of intensity $h$, which favours the rigid spins to be aligned in the direction of the bias, while the soft spins tend to be aligned in the opposite to the bias direction. Then,
 the probability ${\cal P}(\{s_i,\phi_j\})$
of finding a particular configuration $\{s_i,\phi_j\}$ of the spin variables is given by
\begin{align}
\label{a}
{\cal P}(\{s_i,\phi_j\}) = \frac{\prod_{j=1}^{N_2} P(\phi_j)}{{\cal Z}^{(\text{A,B})}} \exp\left[ - \frac{{\cal H}^{(\text{A,B})}_0[\{s_i,\phi_j\}]}{T} + \frac{h}{T}\left(\sum_{i=1}^{N_1} s_i - \sum_{j=1}^{N_2} \phi_j \right)\right] \,,
\end{align}
 where  ${\cal H}^{(\text{A,B})}_0[\{s_i,\phi_j\}]$ is the Hamiltonian
 in the absence of an external bias.
 We will distinguish between two cases with respect to the coupling constants of the spins: in the case A,  we suppose that the interactions between the soft and the rigid spins, as well as between the rigid-rigid and the soft-soft spins
 are the same, and are described by a positive coupling constant $J$, so that
 \begin{align}
 \label{H}
 {\cal H}^{(\text A)}_0[\{s_i,\phi_j\}] = - \frac{J}{2 N} \left(\sum_{i,j=1;\, i \neq j}^{N_1} s_i s_j + \sum_{i,j=1; \,i \neq j}^{N_2} \phi_i \phi_j + 2 \sum_{i=1}^{N_1} \sum_{j=1}^{N_2} s_i \phi_j \right) \,,
 \end{align}
In the case B, we consider a bit more complicated situation when the interactions between the soft and the
rigid spins are different from the interactions between the soft-soft and the rigid-rigid ones.
In the latter case, we suppose that the interactions between the soft-rigid spins are described by a coupling constant~$I_s$, while the interactions between the rigid-rigid and the soft-soft spins are still described by the coupling constant $J$. We will concentrate here on situations when $I_s > J > 0$.
In the case B, the Hamiltonian
 of a mixture of coupled rigid and soft spins in the absence of an external bias
  is defined by
  \begin{align}
 \label{HB}
 {\cal H}^{(\text B)}_0[\{s_i,\phi_j\}] = - \frac{J}{2 N} \left(\sum_{i,j=1;\, i \neq j}^{N_1} s_i s_j + \sum_{i,j=1;\, i \neq j}^{N_2} \phi_i \phi_j\right) - \frac{I_s}{N} \sum_{i=1}^{N_1} \sum_{j=1}^{N_2} s_i \phi_j  \,.
 \end{align}
Both ${\cal H}^{(\text A)}_0[\{s_i,\phi_j\}]$ and ${\cal H}^{(\text B)}_0[\{s_i,\phi_j\}]$ are the Hamiltonians of some ferromagnets
so that for $h=0$ and for sufficiently strong couplings,
the spins will tend to have the same sign.

 Lastly, ${\cal Z}^{(\text{A,B})}$ in \eqref{a} are the partition functions
\begin{align}
\label{Z}
 {\cal Z}^{(\text{A,B})} = \prod_{i=1}^{N_1} \sum_{s_i = \pm 1} \exp\left(\frac{h}{T} s_i \right) \int^{\infty}_{-\infty} \ldots \int^{\infty}_{-\infty} \left[ \prod_{j=1}^{N_2} \rd\phi_j \, P(\phi_j) \, \exp\left(- \frac{h}{T} \phi_j \right) \right] \,
 \exp\left\{ - \frac{{\cal H}^{(\text{A,B})}_0[\{s_i,\phi_j\}]}{T} \right\} \,.
\end{align}
 Note that ${\cal Z}^{(\text{A,B})}$ depend on $T$, $J$, $I_s$ and $h$ only in the combination
$J/T$, $I_s/T$ and $h/T$, so that
 we may safely set $T =1$ supposing that the dependence on the temperature is adsorbed in $J$, $I_s$ and $h$ (or, in other words, that $J$, $I_s$ and $h$ are measured in units of $T$).

In what follows we will be interested in the $h$-dependence of two ``order'' parameters:
\begin{align}
\label{ms}
m^{(\text{A,B})}_s  = \frac{1}{N_1} \sum_{i=1}^{N_1} \langle s_i \rangle \,,
\qquad
m^{(\text{A,B})}_{\phi}  = \frac{1}{N_2} \sum_{j=1}^{N_2} \langle \phi_j \rangle \,,
\end{align}
where the angle brackets denote averaging over the whole society, and also of their linear combination
\begin{align}
\label{M}
M^{(\text{A,B})} = \frac{1}{N} \left(\sum_{i=1}^{N_1} \langle s_i \rangle  + \sum_{j=1}^{N_2} \langle \phi_j \rangle\right) = (1 - \rho) m^{(\text{A,B})}_s + \rho m^{(\text{A,B})}_{\phi} \,.
\end{align}
While $m^{(\text{A,B})}_s$ and $m^{(\text{A,B})}_{\phi}$ define the average ``opinion'' of the ordinary voters and of the contrarians, respectively, the parameter $M^{(\text{A,B})}$
determines
the overall outcome of the vote.

\section{General results for the cases A and B}

We proceed with the calculation of the partition function $ {\cal Z}^{(\text{A,B})}$ in \eqref{Z} focusing first on the case A.

\subsection{The case A}

Since our eventual goal is  the
calculation of the order parameters,
it is expedient to study a little bit
more general object rather than
$ {\cal Z}^{(\text A)} $ itself, letting the biases acting on $s_i$ and $\phi_j$ be different.
To this end, we consider an auxiliary partition function
\begin{align}
\label{Z1}
 {\cal Z}^{(\text A)}(h_1,h_2) = \prod_{i=1}^{N_1} \sum_{s_i = \pm 1} \exp\left(\frac{h_1}{T} s_i \right) \int^{\infty}_{-\infty} \ldots \int^{\infty}_{-\infty} \left[ \prod_{j=1}^{N_2} \rd\phi_j \, P(\phi_j) \, \exp\left(- \frac{h_2}{T} \phi_j \right) \right] \,
 \exp\left\{ - \frac{{\cal H}^{(\text A)}_0[\{s_i,\phi_j\}]}{T} \right\} \,,
\end{align}
which reduces to the expression in \eqref{Z} for $h_1=h_2=h$. Once $ {\cal Z}^{(\text A)}(h_1,h_2)$ is known, the order parameters
can be simply obtained by differentiation of
$ {\cal Z}^{(\text A)}(h_1,h_2)$
as
\begin{align}
m^{(\text A)}_s = \left. \frac{1}{N_1} \frac{\rd}{\rd h_1} \ln  {\cal Z}^{(\text A)}(h_1,h_2)\right|_{h_1=h_2=h},
\qquad m^{(\text A)}_{\phi} = - \left. \frac{1}{N_2} \frac{\rd}{\rd h_2} \ln  {\cal Z}^{(\text A)}(h_1,h_2)\right|_{h_1=h_2=h} .
\end{align}
Now,
we formally rewrite the Hamiltonian in \eqref{H} as
\begin{align}
 {\cal H}^{(\text A)}_0[\{s_i,\phi_j\}] = - \frac{J}{2 N} \left[\left(\sum_{i=1}^{N_1} s_i + \sum_{j=1}^{N_2} \phi_j\right)^2 - N_1 - \sum_{j=1}^{N_2} \phi^2_j \right] \,,
\end{align}
where we used the condition that $s_i^2 \equiv 1$.
Taking advantage of the integral identity
\begin{align}
\exp\left(\frac{b^2}{4 a}\right) = \sqrt{\frac{a}{\pi}} \int^{\infty}_{-\infty} \rd x \exp\left(- a x^2 + b x\right), \qquad a > 0 ,
\end{align}
we cast the auxiliary
partition function in \eqref{Z1} into a factorised, with respect to $s_i$ and $\phi_j$, form, and
 perform straightforwardly averaging over the spin variables.
In doing so, we get
\begin{align}
\label{z}
 {\cal Z}^{(\text A)}(h_1,h_2) = \sqrt{\frac{N}{2 \pi J}} \exp\left[- \frac{J}{2} (1- \rho) + N (1 - \rho) \ln 2 \right] \int^{\infty}_{-\infty} \rd x \exp\left[- N F_{h_1,h_2}(x)\right] \equiv \exp\left[- N f(h_1,h_2)\right],
\end{align}
in which the ``free energy density'' $F(x)$ is defined explicitly by
\begin{align}
\label{F}
F_{h_1,h_2}(x)  = \frac{x^2}{2 J} - (1 - \rho) \ln \cosh\left(x+h_1\right) - \rho \ln\left\{\int^{\infty}_{-\infty} \rd\phi P(\phi) \exp\left[(x-h_2) \phi - \frac{J}{2 N} \phi^2\right]\right\} \,,
\end{align}
and $f(h_1,h_2)$ in the right-hand-side of \eqref{z} is the desired free energy of an auxiliary
 system with fields $h_1$ and $h_2$ acting on the spin variables $s_i$ and $\phi_j$, respectively.

Correspondingly, the order parameters $m^{(\text A)}_s$ and $m^{(\text A)}_{\phi}$ in \eqref{ms}  can be written as
\begin{align}
\label{ms1}
&m^{(\text A)}_s = \dfrac{\int^{\infty}_{-\infty} \rd x \tanh\left(x+ h\right) \exp\left[- N F_{h_1=h_2=h}(x)\right]}{\int^{\infty}_{-\infty} \rd x \exp\left[- N F_{h_1=h_2=h}(x)\right]}\,, \qquad
m^{(\text A)}_{\phi} =  \dfrac{\int^{\infty}_{-\infty} \rd x {\cal L}^{(\text A)}(x) \exp\left[- N F_{h_1=h_2=h}(x)\right]}{\int^{\infty}_{-\infty} \rd x \exp\left[- N F_{h_1=h_2=h}(x)\right]} \,,
\end{align}
with
\begin{align}
\label{LL}
{\cal L}^{(\text A)}(x) = \dfrac{\int^{\infty}_{-\infty} \phi P(\phi) \rd\phi \exp\big[(x - h) \phi - \frac{J}{2 N} \phi^2\big]}{\int^{\infty}_{-\infty}  P(\phi) \rd\phi \exp\big[(x - h) \phi - \frac{J}{2 N} \phi^2\big]} \,.
\end{align}

\begin{figure}[!t]
\centering
\begin{minipage}[b]{0.33\linewidth}
\includegraphics[width = 1. \textwidth]{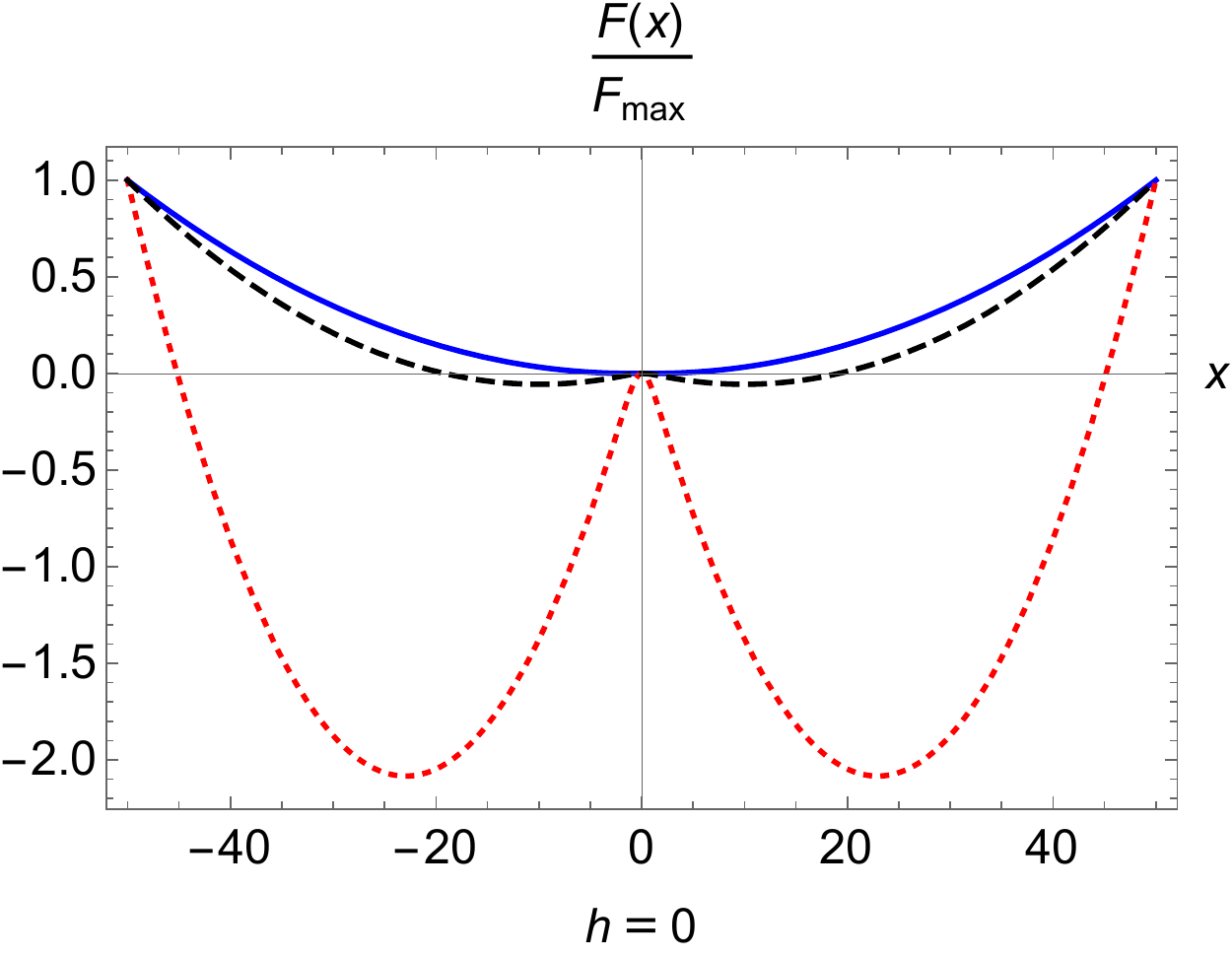}
\end{minipage}
\begin{minipage}[b]{0.33\linewidth}
  \includegraphics[width = 1. \textwidth]{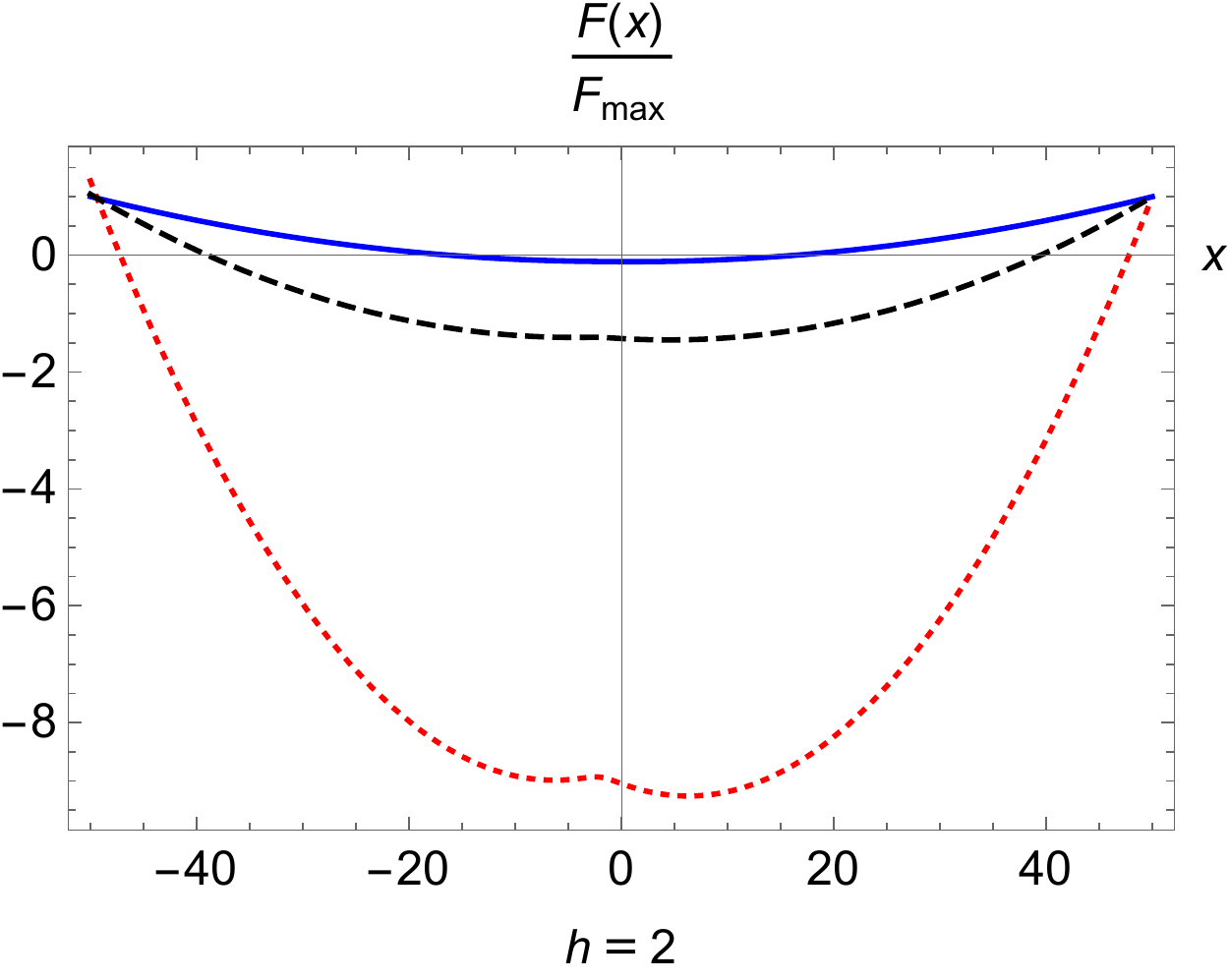}
\end{minipage}
\begin{minipage}[b]{0.33\linewidth}
\includegraphics[width = 1. \textwidth]{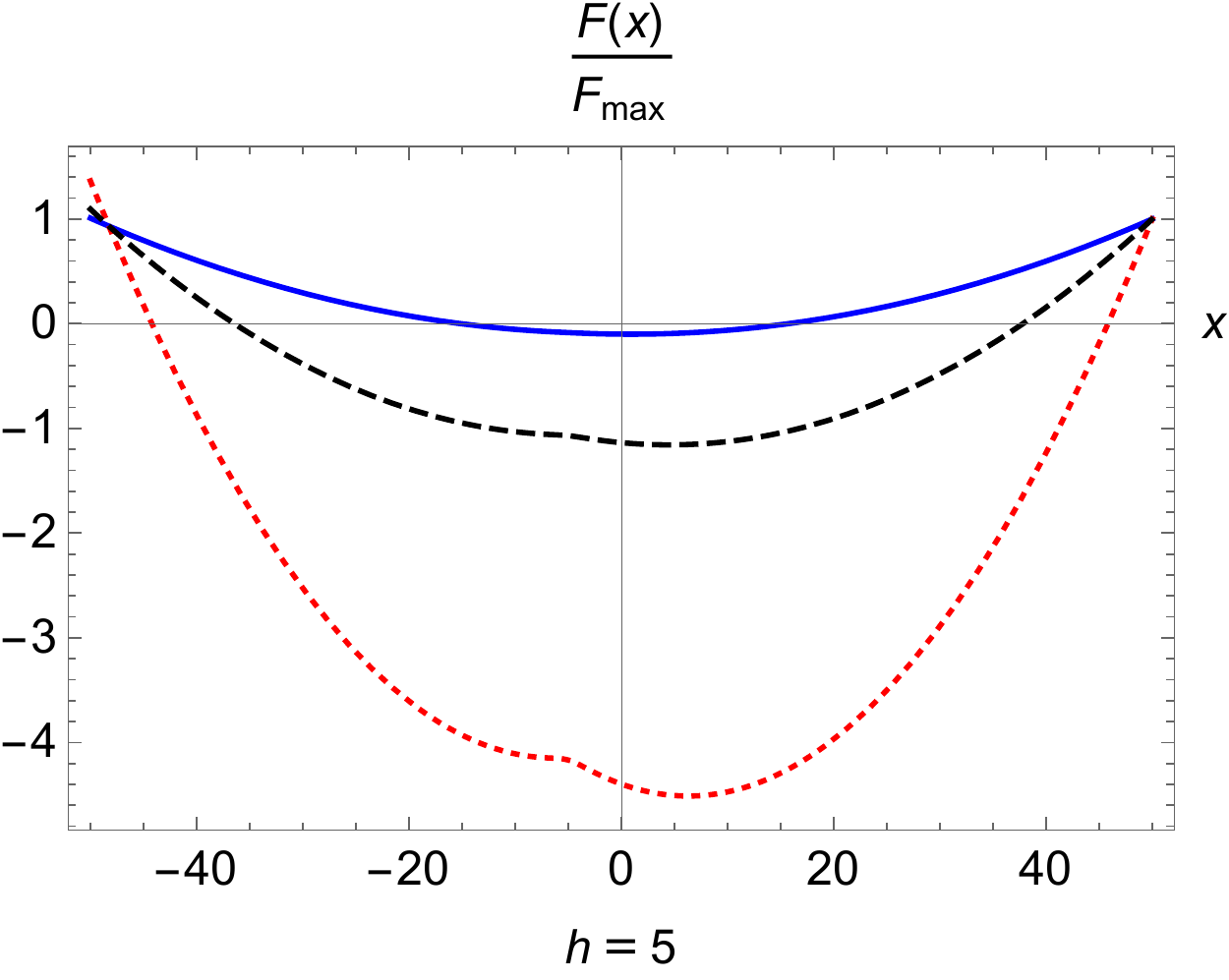}
\end{minipage}
\caption{(Color online) Function $F(x)$ in \eqref{F} (for convenience divided by its maximal value attained at $x = 50$)
for $h_1=h_2=h$, $\rho =0.1$ and $N = 10^3$,
and $P(\phi)$ defined in \eqref{P} with $\sigma_{\phi} = 1$.
Solid curves correspond to $J = 1$, dashed --- to $J = 5$ and dotted --- to $J = 7$.
 \label{FIG1}
}
\end{figure}

One notices next that for $x \to \pm \infty$, the dominant contribution to $F(x)$
comes from the first term in the right-hand-side of \eqref{F},
which diverges in proportion to $x^2$. This signifies
that $F(x)$ has at least one minimum.
More thorough analysis
shows that $F(x)$ has
a single minimum for sufficiently small values of the coupling constant $J$. On the other hand,
for larger values of $J$, $F(x)$ attains a double-well shape. For a particular example of the distribution of the soft spin variables in \eqref{P}, the integral over $\rd\phi$ in the right-hand-side of \eqref{F} can be performed exactly, which permits us to
 illustrate the typical behavior of $F(x)$ as a function of $x$ in different regimes (see figure~\ref{FIG1}).

We focus on the limit of large $N$, $N \gg 1$, when the integrals in \eqref{z} become concentrated in the small vicinity
 of the minimum $x = x^*$ [or two minima, when $F(x)$ has a double-well form]
of $F(x)$, defined by
\begin{align}
\frac{\rd F_{h_1=h_2=h}(x^*)}{\rd x^*} = 0 \,.
\end{align}
Differentiating \eqref{F}, we find the following transcendental equation, which defines $x^*$ implicitly,
\begin{align}
\label{eq}
x^* = J \left\{ (1-\rho) \tanh(x^* + h) + \rho \, \dfrac{\int^{\infty}_{-\infty} \phi\, \rd\phi P(\phi) \exp\big[(x^* - h) \phi - \frac{J}{2 N} \phi^2\big]}{\int^{\infty}_{-\infty} \rd\phi P(\phi) \exp\big[(x^* - h) \phi - \frac{J}{2 N} \phi^2\big]}
\right\} \,.
\end{align}
Further on, performing the integrals in \eqref{z} we find that $m^{(\text A)}_s$ and $m^{(\text A)}_{\phi}$ obey
\begin{align}
\label{ms2}
m^{(\text A)}_s = \tanh(x^* + h) \,,
\qquad m^{(\text A)}_{\phi} = {\cal L}^{(\text A)}(x^*) \,.
\end{align}
Lastly, comparing \eqref{eq} with $m^{(\text A)}_s$ and $m^{(\text A)}_{\phi}$ defined in \eqref{ms2}
and the definition of the order parameter $M^{(\text A)}$ in \eqref{M},
we conclude that $x^*$ has quite a lucid physical meaning: indeed, $x^* = J M^{(\text A)}$ and hence,
$x^*$ defines the outcome of the vote multiplied by $J$.

\subsection{The case B}

The derivation of transcendental equations, which  implicitly define the order parameters $m^{(\text B)}_s$ and $m^{(\text B)}_{\phi}$ is only slightly more involved. Since it follows essentially the same line of thought, we present these equations here without a derivation. We have that in the case B:
\begin{align}
\label{msB}
m^{(\text B)}_s = \tanh\left[J (1-\rho) m^{(\text B)}_s   + I_s \rho m^{(\text B)}_{\phi} + h\right] \,,
\qquad m^{(\text B)}_{\phi} =  {\cal L}^{(\text B)}\left(m_s^{(\text B)},m_{\phi}^{(\text B)},h\right) \,,
\end{align}
where
\begin{align}
\label{AAA}
{\cal L}^{(\text B)}\left(m_s^{(\text B)},m_{\phi}^{(\text B)},h\right) = \dfrac{\int^{\infty}_{-\infty} \phi P(\phi) \rd\phi \exp\left\{\left[J \rho m_{\phi}^{(\text B)} + I_s (1 - \rho) m_s^{(\text B)} - h \right] \phi \right\}}{\int^{\infty}_{-\infty}  P(\phi) \rd\phi \exp\left\{\left[J \rho m_{\phi}^{(\text B)} + I_s (1 - \rho) m_s^{(\text B)} - h \right] \phi\right\}} \,.
\end{align}
The expressions in \eqref{msB} reduce to the ones in \eqref{ms2} for $J = I_s$.
Below we will discuss the behavior of the order parameters in the cases A and B for the distribution $P(\phi)$ in \eqref{P}.

\section{Results for the distribution in \eqref{P}. Case A}

In the particular case when the distribution $P(\phi)$ is given by \eqref{P},
${\cal L}(x^*)$ in the right-hand-side of~\eqref{LL}  can be calculated exactly.
In the leading in the limit $N \to \infty$ order, we have
\begin{align}
{\cal L}(x^*) = (x^* - h) \sigma_{\phi} + \tanh(x^* - h) \,,
\end{align}
so that \eqref{eq} becomes
\begin{align}
\label{MM}
x^* = J \left[ (1 - \rho) \tanh(x^* + h) + \rho  (x^* - h) \sigma_{\phi} + \rho \tanh(x^* - h)
\right] \,.
\end{align}
Two latter coupled transcendental equations, together with \eqref{ms2},  implicitly define the order
parameters in the case A.

These equations, which have multivalued solutions for sufficiently large values of $J$, are quite difficult to study analytically in the whole parameter range (numerical solution is possible, of course).
Nonetheless, they permit us to get some general understanding of the
behavior of the order parameters in the limiting cases of small and large $h$. We consider first
the limit $h \ll 1$, which corresponds to the limit
of a  \textit{linear response} of a system to an external bias. In this limit, for sufficiently small $J$, we have that
 in the leading in $h$ order the order parameter $M^{(\text A)}$ obeys
\begin{align}
\label{770}
M^{(\text A)} =  \frac{\left[1 - \rho (2 + \sigma_{\phi})\right]}{1 - J (1 + \rho \sigma_{\phi})} \, h \,.
\end{align}
Since the prefactor in this dependence is positive definite
for sufficiently small $J$ and $\rho$,  the order parameter is an increasing function of $h$ for small $h$. Further on, we find that in this limit, the order parameter of the soft and of the spins obey
\begin{align}
\label{771}
m^{(\text A)}_{\phi} = - \frac{1 - 2 J (1 - \rho)}{1 - J (1 + \rho \sigma_{\phi})} \, (1 + \sigma_{\phi}) \, h \,,
\qquad m^{(\text A)}_s = \frac{1 - 2  J \rho (1 + \sigma_{\phi})}{1 - J (1 + \rho \sigma_{\phi})} \, h \,.
\end{align}
We notice that for sufficiently small $J$, the order parameter
$m^{(\text A)}_{\phi}$ is a decreasing function of $h$,
while $m^{(\text A)}_s$ increases with an increase of $h$.

Within the opposite
limit $h \to \infty$,
we find from \eqref{MM} that in the leading in $h$ order
\begin{align}
\label{hbigM}
M^{(\text A)} = - \frac{\rho \sigma_{\phi}}{1 - J \rho \sigma_{\phi}} \, h \,,
\end{align}
implying that $M^{(\text A)} \to - \infty$ as $h \to \infty$.
This means, in turn, that  $M^{(\text A)}$ is a non-monotonous function of $h$ and that
 for a strong (excessive) bias, the
outcome of the vote is always against the preferential
candidate.

Regarding the behavior of two other order parameters, we notice that since
$(x^* + h)$
is always positive definite (for $J \rho \sigma_{\phi} < 1$), the order parameter of the rigid spins $m_s^{(\text A)} \to 1$ as $h \to \infty$, which means that the whole subgroup of ordinary voters is in favor of the preferential candidate. On the contrary, the order parameter of the contrarians behaves, in the leading in $h$ order, as
\begin{align}
\label{hbigmphi}
m^{(\text A)}_{\phi} = - \frac{\sigma_{\phi}}{1 - J \rho \sigma_{\phi}} \, h \,,
\end{align}
i.e., is negative and is  growing indefinitely  by an absolute value with $h$ so that
the order parameter $m^{(\text A)}_{\phi}$ for sufficiently small $J$ is always a monotonously decreasing
function of $h$.
In conclusion, we have that in the case A, an excessive bias breaks the ties within the community,
letting the ordinary voters to have an opinion in favor of the preferential candidate,
and the contrarians to have an opinion against this candidate. However,
overall at large $h$, the contrarians dominate by increasing the strength of their opinion.

We proceed with the numerical analysis,
which permits us to have a broader look at the behavior of such a
 mixture of mutually-interacting rigid and soft spins.
 In particular, it permits us to determine the order parameters at arbitrary, not necessarily small or big values of $h$,
 to consider the situations with arbitrary
  values of the coupling
  $J$ and also to analyse the effect of the initial configuration.
  In our numerical Monte Carlo simulations, we first fix an initial configuration of the spin variables, (e.g., stipulating
  that initially all spins are pointing downwards, or upwards, or are in some mixed configuration),
  fix $J$ and $h$,   and then let the system equilibrate
  via standard Metropolis algorithm for a sufficiently
  long time (typically, for $10^8$ time steps)
  until it reaches an equilibrium configuration, which may, in general, be different from the initial one.
  The purpose of doing so is as follows: for sufficiently large $J$ and for $h = 0$, the ferromagnetic
  system under study acquires a spontaneous
  ``magnetization''
  so that the order parameters $m_s^{(\text A)}(h=0)$ and $m_{\phi}^{(\text A)}(h=0)$ are no longer equal to zero, unlike
  the case of small $J$ when both $m_s^{(\text A)}(h=0) = m_{\phi}^{(\text A)}(h=0) = 0$. The order parameter $m_s^{(\text A)}(h=0)$, for example, attains two values $m_s^{(\text A)}(h=0) = \pm m_s$, where positive and negative values are chosen by the system
  with equal probability. In other words, for $h=0$, it is equally probable that the whole system will have a negative or a positive magnetization.
  By choosing the initial state, i.e., by forcing the spins to point in some direction,
  we break the symmetry between the negative and positive value
  ``helping'' the system to arrive into the state with a prescribed sign of magnetization.

\begin{figure}[!t]
\includegraphics[width = 1. \textwidth]{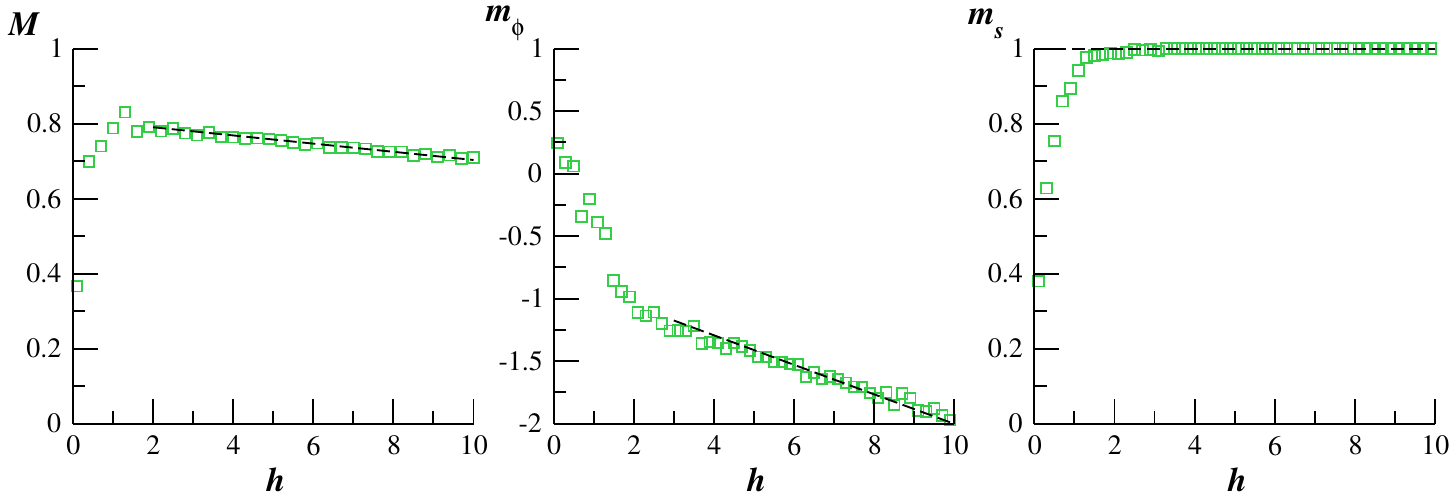}
\centering
\caption{(Color online) Order parameters as functions of the external bias $h$ for
  $J=0.75$, $\sigma_\phi=0.1$, $\rho=0.1$ and $N=1111$. Initially all spins (both rigid and soft ones) are all
   set equal to $+1$ and the system is let to equilibrate for $10^8$ time steps at fixed $J$ and $h$.
   The dashed lines stand for the corresponding asymptotic
  results in \eqref{hbigM} (left-hand panel), \eqref{hbigmphi} (central panel) and $m_s^{(\text A)} = 1$ (right-hand panel).
 \label{FIG2}
}
\end{figure}

  In figures~\ref{FIG2}, we first depict
   the order parameters as functions of $h$ for a small value of $J$, $J = 0.75$ and for
   an initial
   configuration in which all the spins are pointing upwards and which is let to equilibrate at a fixed $J$ and $h$    for $10^8$ time steps.    We observe that for such a value of $J$,
   upon an increase of $h$, the order parameter of the rigid spins
  grows monotonously with $h$
  and quite rapidly, already for $h \approx 2$, attains
   the limiting value $m_s^{(\text A)} = 1$.
   The order parameter $m_{\phi}^{(\text A)}$ of the soft spins, on the contrary, is a monotonously decreasing function of $h$. This property also
   quite rapidly, already for $h \approx 2$, converges to the asymptotic form in \eqref{hbigmphi}.
   The combined effect of $m_{s}^{(\text A)}$ and $m_{\phi}^{(\text A)}$ is such that the  order parameter $M^{(\text A)}$ first increases,
   being dominated by the rigid spins, passes through a maximal value,
   attained at $h \approx 2$, and then gradually decreases with $h$, becoming dominated
    by the soft spins. Overall, we observe a very good agreement between the numerical results and our predictions for the asymptotic large-$h$ behavior.

 \begin{figure}[!b]
\includegraphics[width = 1. \textwidth]{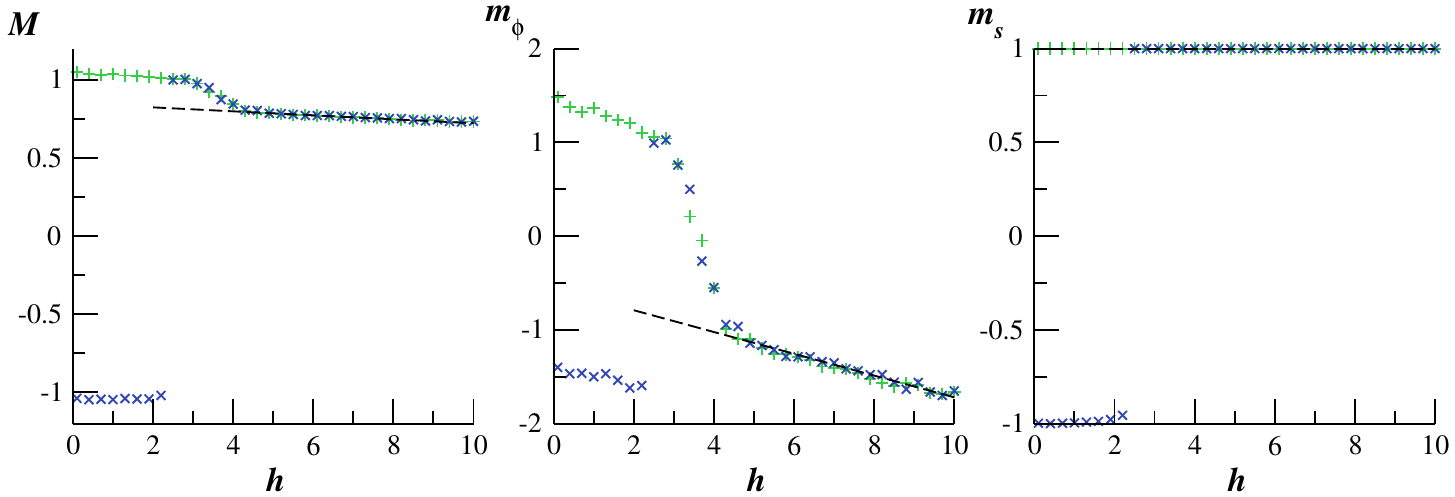}
\centering
\caption{(Color online) Order parameters as functions of the external bias $h$ for
  $J=4$, $\sigma_\phi=0.1$, $\rho=0.1$ with $N=1111$. Symbols correspond to two different
  initial states: $\{s_i\} = \{\phi_j\}=+1$ (pluses) and $\{s_i\} =
  \{\phi_j\}=-1$ (crosses), which are subsequently let to equilibrate for $10^8$ time steps.
   The dashed lines stand for the corresponding asymptotic
  results in \eqref{hbigM} (left-hand panel), \eqref{hbigmphi} (central panel) and $m_s^{(\text A)} = 1$ (right-hand panel).
 \label{FIG3}
}
\end{figure}

 In figures~\ref{FIG3} we plot the order parameters as functions of $h$ for a sufficiently big value of $J$, $J= 4$, (such that the system may acquire a spontaneous magnetization at $h=0$), and also for two different initial
 configurations of spins. We start our discussion with the case when the
 initial configuration is most favorable for the preferential candidate --- when all the rigid and the soft spins point upwards (curves depicted by pluses in figures~\ref{FIG3}) and then are let to equilibrate within $10^{10}$ time steps. In this situation, the order parameter $m_s^{(\text A)}$ of the rigid spins does not show any appreciable variation with $h$ and stays constant and equal to $1$ for any $h$. The order parameter $m_{\phi}^{(\text A)}$ appears to be a monotonously decreasing function of $h$ and quite rapidly, already
 at $h \approx 4$, starts to converge to the asymptotic form  in~\eqref{hbigmphi}.  Overall, the order parameter $M^{(\text A)}$ is also a monotonously decreasing function of $h$, which means that prompting such a community to vote for a preferential candidate appears to be counter-productive even at very small values of $h$.

 For the initial configuration when all the spins --- the rigid and the soft ones --- point downwards and are let to equilibrate afterwards, the behavior of the order parameters appears to be more interesting and rich,
 and exhibits a seemingly discontinuous variation with the value of the external bias resembling true phase transitions in physical systems.  We observe that here the order parameter $m_s^{(\text A)}$ stays equal to its value at $h = 0$, $m_s^{(\text A)}(h) \approx - 1$, up to a certain threshold value of $h$, (which only slightly exceeds $h =2$),
 and then rather abruptly jumps  to
 $m_s^{(\text A)}(h) \approx + 1$, when the bias becomes strong enough
 to pay the penalty
 set by the interaction energy with the soft spins.
 The order parameter $m_{\phi}^{(\text A)}$ first slightly
 decreases with an increase of $h$ and then abruptly jumps upwards, being turned over by the interactions with the rigid spins. In a way, one may claim that here the opinion of the contrarians gets inverted by the opinion of the ordinary voters. Further,  $m_{\phi}^{(\text A)}$ monotonously decreases from the peak value and ultimately, for $h \approx 4$, converges to the asymptotic form in \eqref{hbigmphi}.
 The global order parameter
 $M^{(\text A)}$, which represents the combined effect of the rigid and the soft spins, follows mostly the behavior of the order parameter of the soft spins and exhibits a discontinuous jump from  $M^{(\text A)} \approx - 1$ to $M^{(\text A)} \approx + 1$ at $h \approx 2$, and then gradually decreases upon an increase of the external bias. Here, as well, we observe a very good agreement between the numerical data and our predictions for the asymptotic large-$h$ behavior.

\section{Results for the distribution in \eqref{P}. Case B}

In this case, we have that the order parameter $m_s^{(\text B)}$ obeys a transcendental \eqref{msB}, which couples it to the order parameter $m_{\phi}^{(\text B)}$. For the latter, performing the integrals in \eqref{AAA} with the distribution in \eqref{P}, we find the following equation
\begin{align}
\label{mphiBB}
m_{\phi}^{(\text B)} = \sigma_{\phi} \left[J \rho m_{\phi}^{(\text B)} + I_s (1 - \rho) m_s^{(\text B)} - h\right] + \tanh\left[J \rho m_{\phi}^{(\text B)} + I_s (1 - \rho) m_s^{(\text B)} - h\right] \,.
\end{align}
As in the previous section, we start here with the solution of the coupled  transcendental equations \eqref{msB} and  \eqref{mphiBB} in the limits $h \ll 1$ and $h \to \infty$. In the limit of a linear response, we have
\begin{align}
m_s^{(\text B)} = \frac{1 - \rho \left(J + I_s\right) (1 + \sigma_{\phi})}{\left[1 - J (1 + \rho \sigma_{\phi}) +\left(J^2 - I^2_s\right) \rho (1-\rho) (1 + \sigma_{\phi})\right]} \, h \,,
\end{align}
\begin{align}
m_{\phi}^{(\text B)} = - \frac{1 - \left(J + I_s\right) (1 - \rho)}{\left[1 - J (1 + \rho \sigma_{\phi}) +\left(J^2 - I^2_s\right) \rho (1-\rho) (1 + \sigma_{\phi})\right]} (1 + \sigma_{\phi}) h \,,
\end{align}
and
 \begin{align}
M^{(\text B)} = \frac{1 - \rho (2 + \sigma_{\phi})}{\left[1 - J (1 + \rho \sigma_{\phi}) +\left(J^2 - I^2_s\right) \rho (1-\rho) (1 + \sigma_{\phi})\right]}(1 + \sigma_{\phi}) h \,.
\end{align}
These expressions simplify to the ones given by \eqref{770} and \eqref{771} in which we set $J = I_s$ and, in principle, show a very similar behavior. On the contrary,
the behavior in the limit $h \to \infty$ appears to be quite different.
At the same time, the order parameters $m_{\phi}^{(\text B)}$ and  $M^{(\text B)}$ have
 the same linear dependence on $h$ [compare with \eqref{hbigM} and \eqref{hbigmphi}]; that being
\begin{align}
m_{\phi}^{(\text B)} = - \frac{\sigma_{\phi}}{1 - J \rho \sigma_{\phi}} \, h \,,
\qquad M^{(\text B)} = - \frac{\rho \sigma_{\phi}}{1 - J \rho \sigma_{\phi}} \, h \,,
\end{align}
and hence, both tend to $-\infty$ when $h \to \infty$,
the order parameter $m_{s}^{(\text B)}$ exhibits some novel and interesting features.
The point is that, unlike the case A, here the argument in the hyperbolic tangent in  \eqref{msB} can be positive or negative depending on the value of the coupling parameter $I_s$. More specifically, we have that
for $I_s < I_s^{\,\text c}$, where
\begin{align}
I_s^{\,\text c} \approx \frac{1}{\rho \sigma_{\phi}} - J \,,
\end{align}
the order parameter $m_{s}^{(\text B)} \to 1$ as $h \to \infty$, while for $I_s > I_s^{\,\text c}$ the order parameter  $m_{s}^{(\text B)} \to -1$. This means that in the case B, these are the soft spins which may turn over the rigid ones,  present in majority. This is an opposite effect, as compared to  the phenomenon which we observed
for the case A.

\begin{figure}[!b]
\centering
\begin{minipage}[b]{0.41\linewidth}
\includegraphics[width = 0.93 \textwidth]{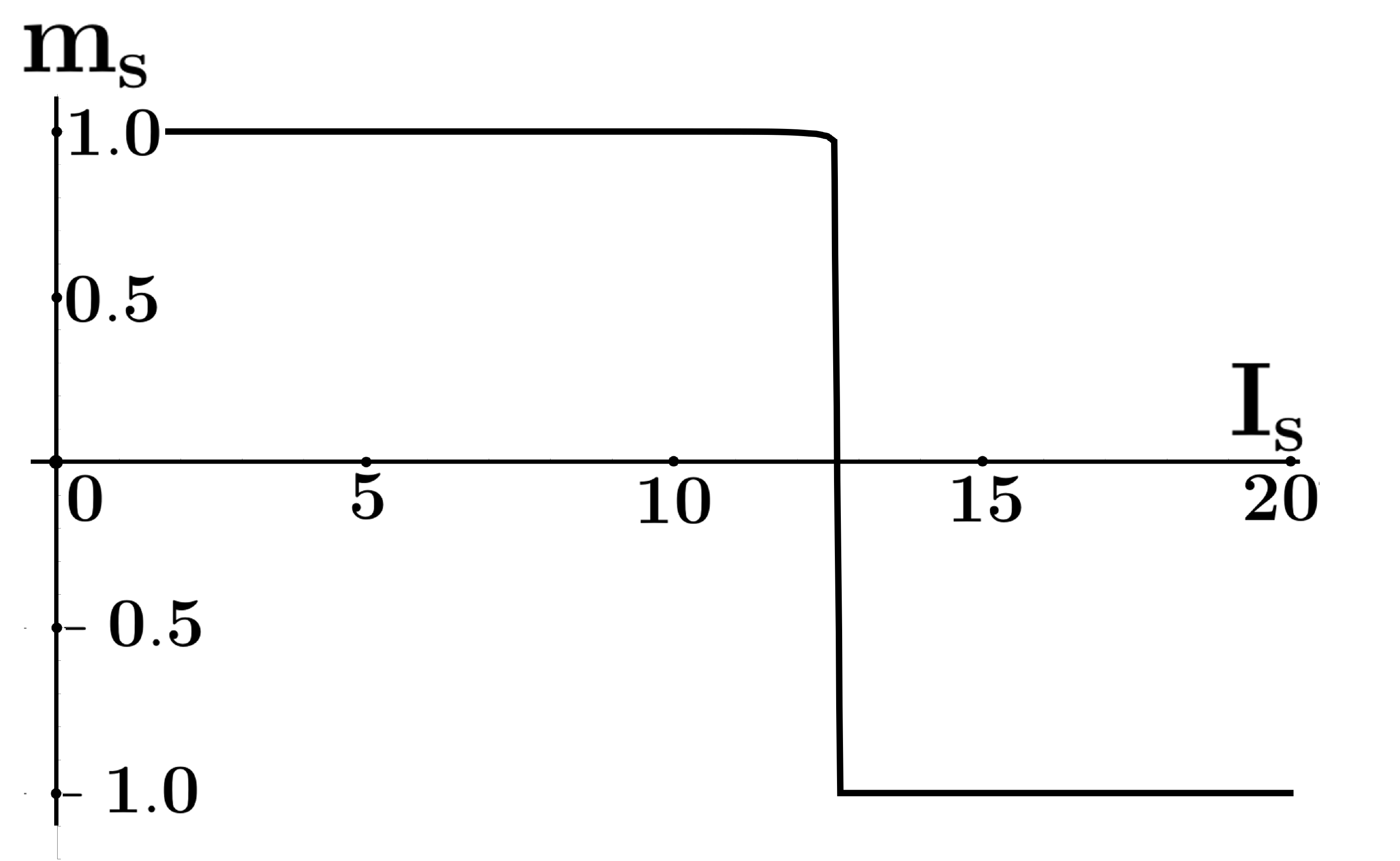}
\end{minipage}
\quad
\begin{minipage}[b]{0.41\linewidth}
  \includegraphics[width = 0.93 \textwidth]{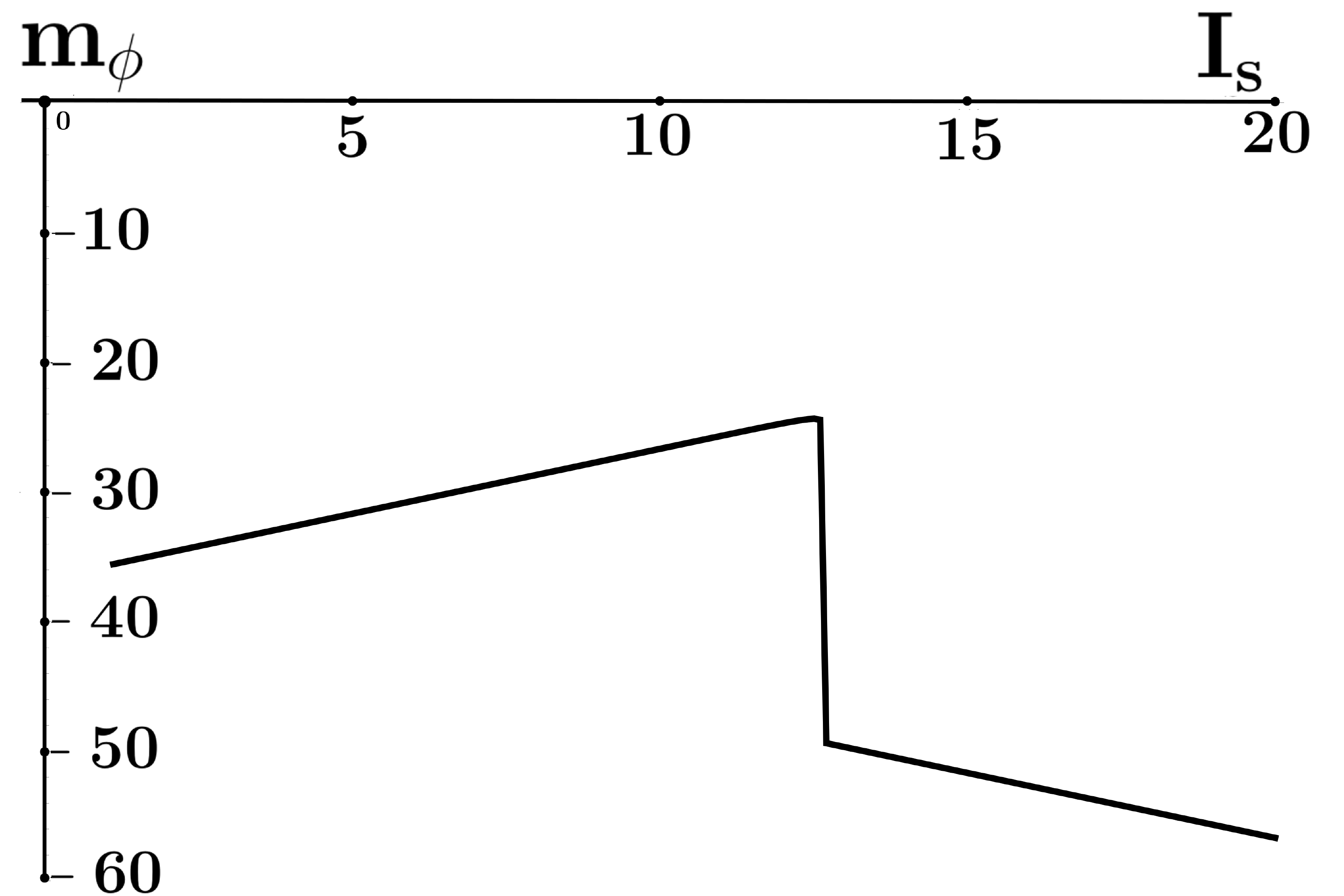}
\end{minipage}
\caption{Numerical solution of \eqref{msB} and \eqref{mphiBB} for $J = 1$, $\sigma_{\phi}=1$, $\rho = 0.1$ and $h = 32$.  Left-hand panel presents $m_{s}^{(\text B)}$, while the right-hand one presents $m_{\phi}^{(\text B)}$ as functions of $I_s$.
 \label{FIG4}
}
\end{figure}
\begin{figure}[!b]
\centering
\begin{minipage}[b]{0.41\linewidth}
\includegraphics[width = 0.93 \textwidth]{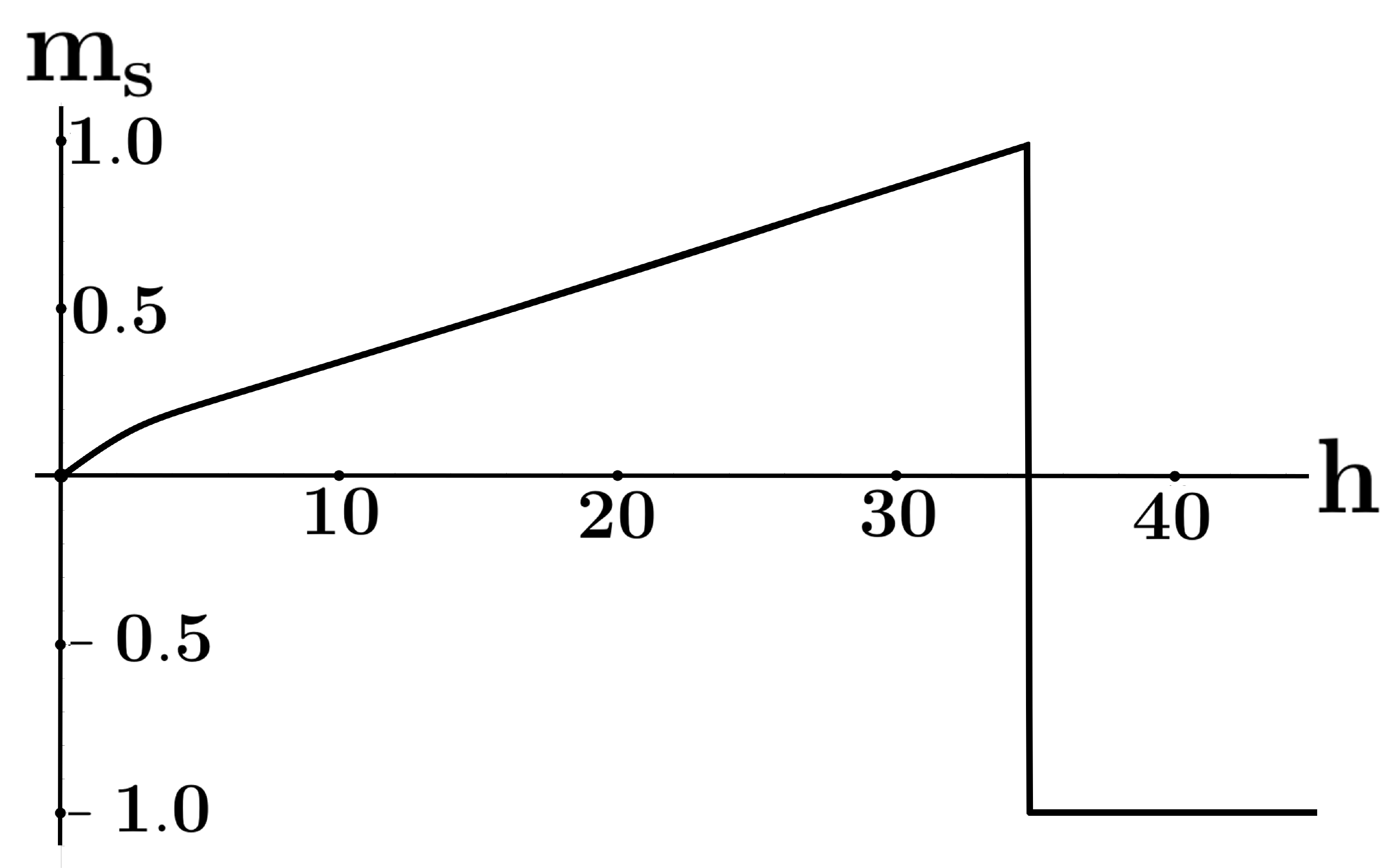}
\end{minipage}
\quad
\begin{minipage}[b]{0.41\linewidth}
  \includegraphics[width = 0.93 \textwidth]{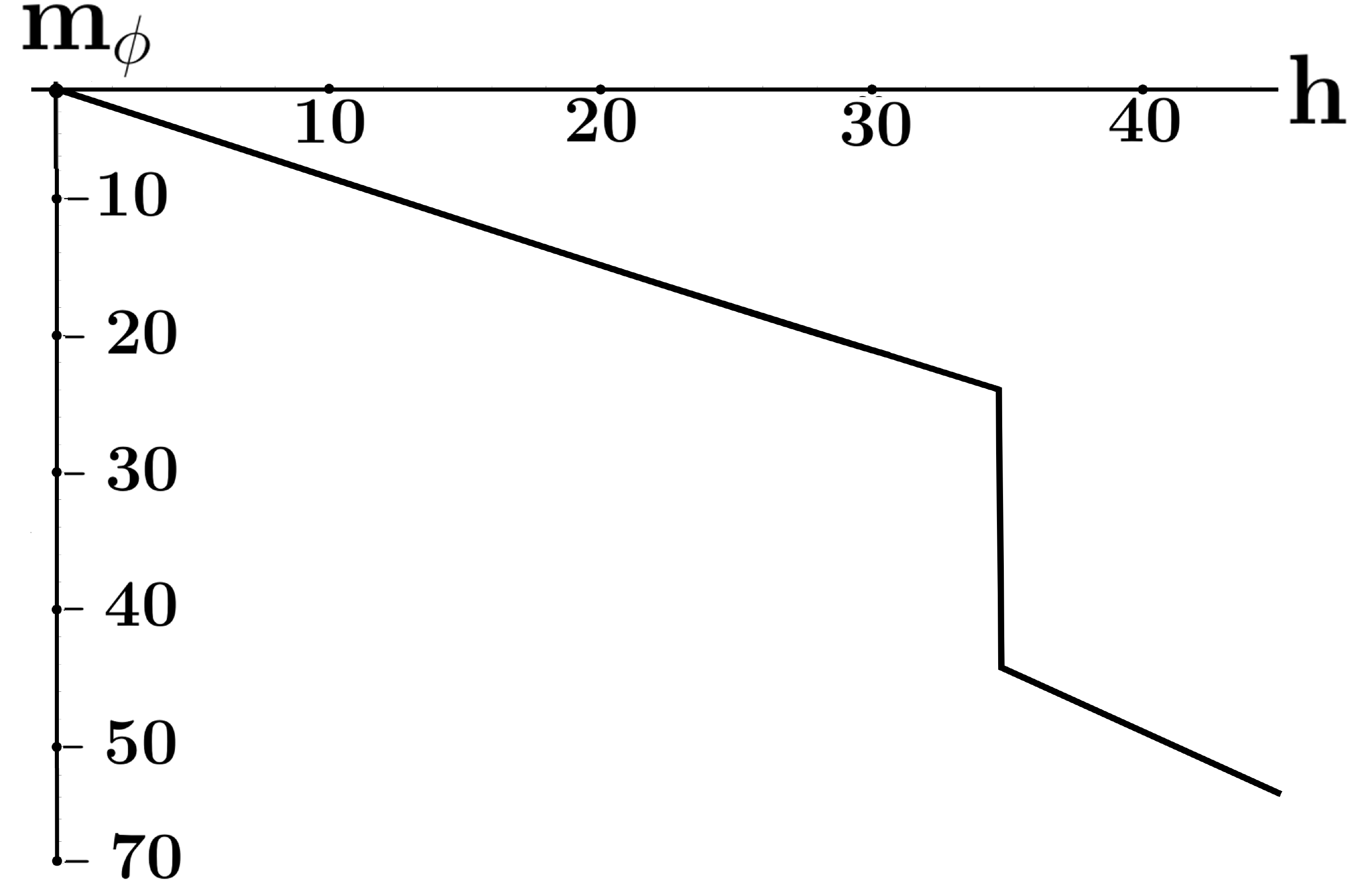}
\end{minipage}
\caption{Numerical solution of \eqref{msB} and \eqref{mphiBB} for $J = 1$, $\sigma_{\phi}=1$, $\rho = 0.1$ and $I_s = 12.5$.  Left-hand panel presents $m_{s}^{(\text B)}$, while the right-hand one presents $m_{\phi}^{(\text B)}$ as functions of $h$.
 \label{FIG5}
}
\end{figure}
\begin{figure}[!b]
\centering
\begin{minipage}[b]{0.41\linewidth}
\includegraphics[width = 0.93 \textwidth]{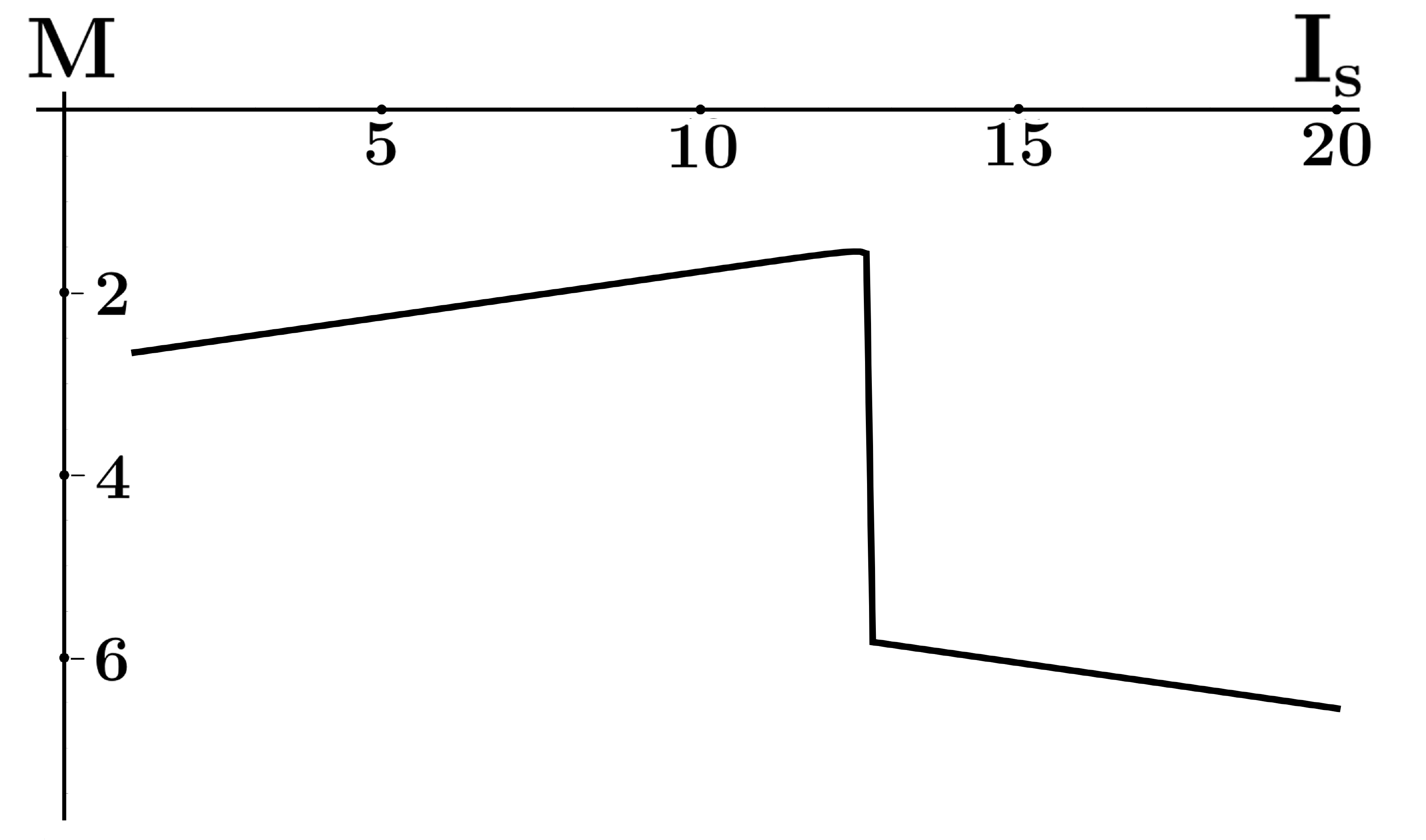}
\end{minipage}
\quad
\begin{minipage}[b]{0.41\linewidth}
  \includegraphics[width = 0.93 \textwidth]{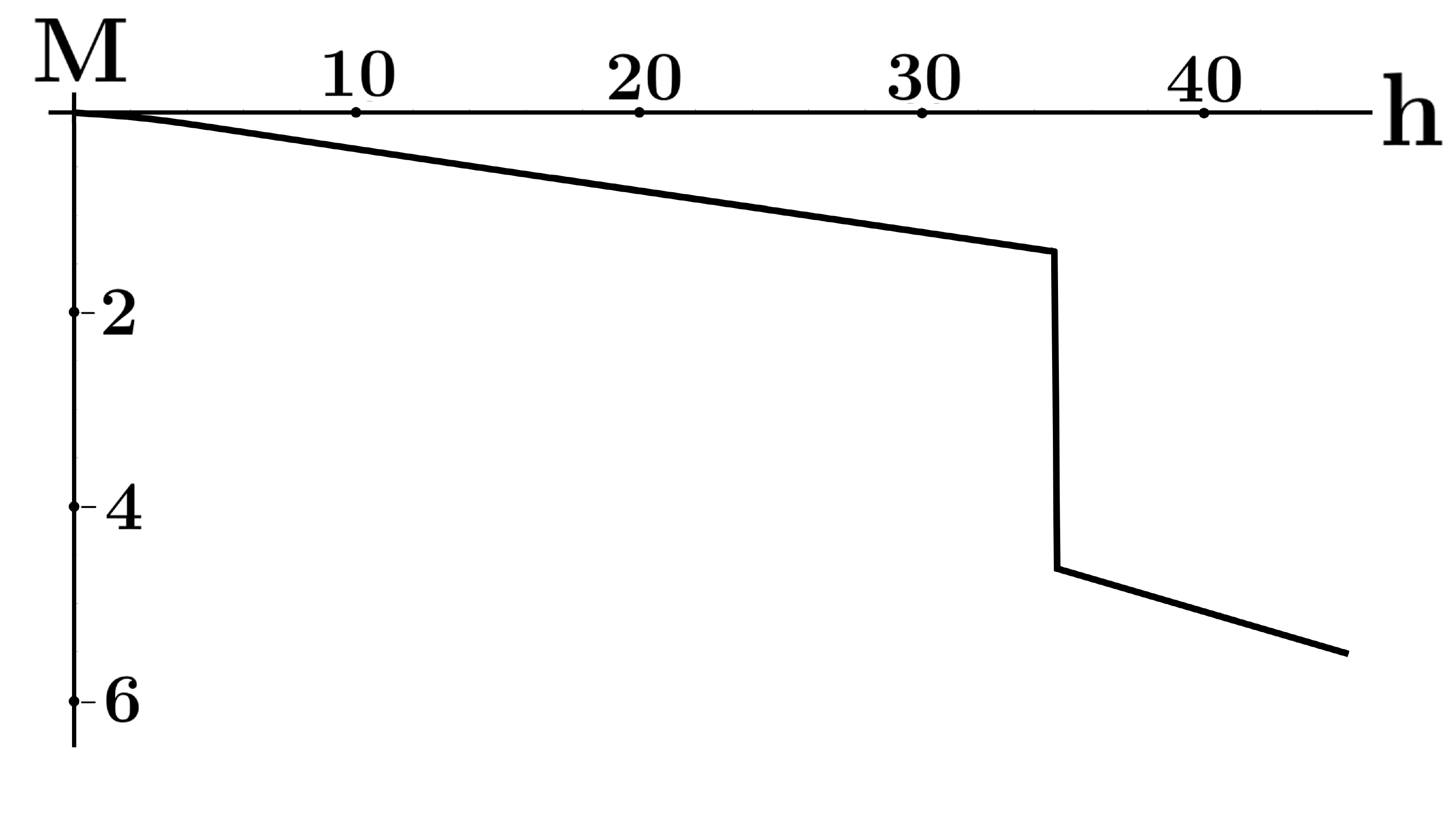}
\end{minipage}
\caption{Numerical solution of \eqref{msB} and \eqref{mphiBB} for $J = 1$, $\sigma_{\phi}=1$, $\rho = 0.1$ and $I_s = 12.5$.  Left-hand panel presents
the global order parameter $M^{(\text B)}$ vs $I_s$ for $h = 32$,
while the right-hand one presents $M^{(\text B)}$ as a function of $h$ for $I_s = 12.5$.
 \label{FIG6}
}
\end{figure}
We illustrate the inversion of the rigid spins by the soft ones
by solving the coupled transcendental equations \eqref{msB} and \eqref{mphiBB} numerically, which also permits us to highlight some interesting features of the behavior at the intermediate values of $h$ and $I_s$. In figures~\ref{FIG4} we depict $m_s^{(\text B)}$ and $m_{\phi}^{(\text B)}$ as functions of $I_s$ for fixed $J=1$ and $h=32$. We observe that $m_{\phi}^{(\text B)}$ is a growing function of $I_s$ for $I_s \leqslant 12.5$. At $I_s = 12.5$, the order parameter abruptly jumps downwards and starts to decrease with a further increase of $I_s$. The downward jump of $m_{\phi}^{(\text B)}$ provokes all the rigid spins to turn over their direction, so that $m_s^{(\text B)}$  discontinuously changes from $m_s^{(\text B)} = 1$ to  $m_s^{(\text B)} = - 1$. Further, in figures~\ref{FIG5} we plot
$m_s^{(\text B)}$ and $m_{\phi}^{(\text B)}$ as functions of $h$ for fixed $J=1$ and $I_s= 12.5$. We notice that $m_{\phi}^{(\text B)}$ is a
decreasing function of $h$ with a discontinuous downward jump at $h \approx 36$. The order parameter  $m_s^{(\text B)}$ first is an increasing
function of $h$, up to the value of $h$ at which $m_{\phi}^{(\text B)}$ jumps down: here, as well,  $m_s^{(\text B)}$ exhibits a discontinuous jump to $-1$ and stays constant further. Lastly,  in figure~\ref{FIG6}, we depict the global order parameter $M^{(\text B)}$ as a function of $I_s$ (left-hand panel) and as a function of $h$ (right-hand panel). We observe that its behavior is entirely dominated by the behavior of the order parameter of the soft spins.

\section*{Conclusions}

To recap, we have studied here a toy model
of opinion formation in a mixed society
consisting of two groups of individuals, interacting all with all across both groups, and
that differ in their response to an external bias
favoring one of two potential candidates in elections. While one group of individuals, present in majority, is characterized by their eagerness to accept the externally favored candidate,
the minority group --- the contrarians --- are antagonistic to the bias and tend to have an opinion
different from the externally imposed one.
Modelling such a society as a system of appropriately defined interacting, effective spin variables,
we arrive at a general conclusion that pushing too strongly for the preferential candidate appears to be counter-productive and
leads to a smaller effect than the one achieved with a modest bias.
We show, as well,
that depending on how the interactions between the two groups and within each of the groups are defined,
 interesting phenomena
of an opinion turn-over can take place, when either of two groups can abruptly change its opinion in favor of the opinion of another group.

\section*{Acknowledgements}

We express our warmest thanks to Dr.
Olesya Mryglod for her interest in this work
and stimulating communications.
C.M.-M. acknowledges a partial
support from the Spanish MICINN grant MTM2015-63914-P.

\ukrainianpart

\title{Негативний відгук на надлишкове відхилення, спричинене змішаною популяцією виборців}
\author[]{В.С. Доценко\refaddr{label1},
        К. Мехія-Монастеріо\refaddr{label2}, Г. Ошанін\refaddr{label1}}
\addresses{
\addr{label1} Лабораторія теоретичної фізики конденсованої матерії, UPMC,
CNRS UMR 7600, університет Сорбонна, 75252 Париж, Франція
\addr{label2} Лабораторія фізичних властивостей, Мадридський технічний університет, 28040 Мадрид, Іспанія
}

\makeukrtitle

\begin{abstract}
Ми вивчаємо результат голосування в популяції виборців, які перебувають під дією зовнішнього зміщення на користь одного з потенційних
кандидатів.
Популяція складається зі звичайних індивідуумів, які становлять більшість і чия думка має тенденцію  до вирівнювання під дією зовнішнього
зміщення, і з деякого числа протилежно настроєних індивідуумів, які завжди є противниками зміщення, але не знаходяться в конфлікті з простими
виборцями.
Виборці взаємодіють між собою як всі зі всіма, намагаючись прийти до спільної думки спільноти як цілого. Ми показуємо, що
для достатньо слабого зовнішнього зміщення, думка звичайних індивідуумів є завжди вирішальною, і результат голосування є в користь
преференційного кандидата.
Навпаки, для надто сильного зміщення протилежно настроєні індивідууми  домінують, приводячи вцілому до негативного відгуку на прикладене
зміщення. Ми також показуємо, що для достатньо сильних взаємодій в межах спільноти, будь-яка з двох підгруп може несподівано
змінити думку іншої групи.

\keywords нелінійний і негативний відгук, зовнішнє зміщення, популяція виборців
\end{abstract}

\end{document}